\documentclass[twocolumn,twoside,superscriptaddress,longbibliography]{revtex4-2}
\usepackage[latin9]{inputenc}
\usepackage{color}
\usepackage{graphicx}
\usepackage{amsmath}
\usepackage{amssymb}
\usepackage[version=3]{mhchem}
\usepackage{float}
\usepackage{soul}
\usepackage{changes}
\usepackage{mathrsfs}
\usepackage{comment}
\usepackage{multirow}
\usepackage{xcolor}
\usepackage{hyperref}

\definecolor{kc}{rgb}{0.6,0,0.6}

\begin{document}

\title{Spin Relaxation and Diffusion in Monolayer 1T'-WTe$_2$ from First-Principles}  
\setcounter{page}{1}   
%\date{\today} 
\author{Junqing Xu\footnotemark[1]} 
\thanks{Current Address: Department of Physics, Hefei University of Technology, Hefei, Anhui, China}
\affiliation{Department of Chemistry and Biochemistry, University of California, Santa Cruz, CA 95064, USA}
\author{Hiroyuki Takenaka\footnotemark[1]} 
\thanks{HT and JX contributed equally} 
\email{Deceased: Hiroyuki Takenaka}
\affiliation{Department of Chemistry and Biochemistry, University of California, Santa Cruz, CA 95064, USA} 
\author{Andrew Grieder} %TBD
\affiliation{Department of Materials Science and Engineering, University of Wisconsin-Madison, WI, 53706, USA}  
\author{Jacopo Simoni}
\affiliation{Department of Materials Science and Engineering, University of Wisconsin-Madison, WI, 53706, USA}
\author{Ravishankar Sundararaman}
\email{sundar@rpi.edu}  
\affiliation{Department of Materials Science and Engineering, Rensselaer Polytechnic Institute, 110 8th Street, Troy, New York 12180, USA}  
\author{Yuan Ping}
\email{yping3@wisc.edu}
\affiliation{Department of Materials Science and Engineering, University of Wisconsin-Madison, WI, 53706, USA}
\affiliation{Department of Physics, University of Wisconsin - Madison, WI, 53706, USA}
\affiliation{Department of Chemistry, University of Wisconsin - Madison, WI, 53706, USA}

\begin{abstract}
%\textcolor{red}{YP: this abstract is too detailed. You need to make it more general that gives broad impact. focus less on number ; but physics and application implications} 
Understanding spin relaxation in topological systems such as quantum spin-hall (QSH) insulator is critical for realizing coherent transport at high temperature. WTe$_{2}$, known as a QSH insulator with a high transition temperature of 100K, is an important test-bed of unveiling spin relaxation mechanism in topological materials. In this work, we employ our recently-developed
\emph{ab initio} density-matrix dynamics approach to investigate spin relaxation mechanism, and calculate spin lifetime and diffusion length of monolayer 1T'-WTe$_{2}$,  at finite temperature under an external electric field. We found the spin lifetime of electrons have the largest anisotropy when measuring along the canted-spin-texture direction. Moreover, we found an opposite trend between spin and carrier relaxation against applied electric field. Most importantly, the relaxation mechanism under intermediate electric field around 1V/nm can not be explained by either Eillot-Yafet or Dyakonov-Perel models, which highlights the generality of our \emph{ab initio} density-matrix framework. We then proposed analytical models to explain its mechanism and compare well with \emph{ab initio} results at small and large electric field. %The proposed analytical models can explain the trends of spin relaxation mechanism at the limit of small or large electric fields. 
We predict that spin lifetime and spin diffusion length of bulk-state
electrons are $\sim$1 ps and $\sim$30 nm at room temperature respectively, suggesting its promise for spintronic applications.

\end{abstract}

\maketitle

\section*{Introduction}

Realization of coherent transport at room temperature is the Holy Grail of modern electronics. Quantum spin hall insulator (QSHI) enables coherent spin transport at its edge states, protected against scatterings. However, such coherent transport can be only realized at a few Kelvins in most known QSHI, which prohibits its practical applications in spintronics. 1T' WTe$_{2}$, known as a QSHI~\citep{Tang2017-ja}, set a high transition temperature about 100 K~\citep{Fei2017-nw}. Although this transition temperature was mostly argued related to their quasiparticle band gaps, recent ab-initio theory shows a much higher transition temperature could be derived from its band gap, with negligible effect from 
finite temperature state renormalization~\cite{marrazzo2022thermal}.  The possible reason  for the lower observed transition temperature than theoretical prediction was attributed to the scattering from the bulk states, which mix with edge states, leading to spin relaxation and incoherent spin transport. Understanding the dominant relaxation mechanism then suppressing such processes will further raise the transition temperature, approaching  room temperature QSHI. 
%Therefore, answer the key questions of what is the dominant spin relaxation mechanism in monolayer 1T' WTe$_{2}$,  and how it responds to external electric field as an effective tuning nob. 

Experimentally, the related observables are spin relaxation time ($\tau_s$) and spin diffusion length ($l_s$), %which are also critical parameters for spintronics applications, 
required to
be sufficiently long for stable detection and manipulation of spin in applications.
In 1T' WTe$_2$, values of large variations were reported - from 0.2 ps to 1.2 ns for $\tau_{s}$ in Refs. \citenum{chen2020anisotropic,lee2019spin,chen2022observation,seifert2019plane,wang2018room}, 
and from 8 nm to 2.2 $\mu$m for $l_{s}$ in Refs. \citenum{zhao2020observation,zhang2017tunable,song2020coexistence} at room temperature. 
%\textcolor{blue}{at what temperatures?}
Therefore, accurate theoretical prediction of spin relaxation time and diffusion length will serve as an important reference for interpreting experimental results. More importantly, determining the dominant spin relaxation mechanism for better designing high temperature QSHI is the most critical question here, which is the focus of this work.

Previously, $l_{s}$ of monolayer WTe$_{2}$ (ML-WTe$_{2}$) at zero
temperature was simulated using tight-binding Hamiltonian and Landauer-B{\"u}ttiker
formalism\cite{vila2021low}. In their simulations, the elastic scattering
was considered by adding Anderson disorder to mimic the inhomogeneity or disorder in the samples. %\textcolor{red}{what did they obtain for ls?} 
However, explicit scattering through electron-phonon coupling, which is critical for room temperature spin relaxation has not been studied.
Furthermore, neither the spin relaxation mechanism nor its lifetime were studied. 
In this work, we simulate
$\tau_{s}$ of 1T' ML-WTe$_{2}$ using our first-principles real-time density-matrix (FPDM) approach\cite{xu2020spin,xu2021ab,xu2021giant,xu2022substrate,xu2022spin,habib2022electric}, 
with self-consistent spin-orbit coupling (SOC)
and quantum
description of the electron-phonon (e-ph) scatterings. We compute $l_s$ through its  relation with spin lifetime and diffusion 
coefficient ($l_{s}=\sqrt{D\tau_{s}}$)~\cite{vzutic2004spintronics, wu2010spin}. Through our calculations, we answered the key questions - the dominant spin relaxation mechanism in monolayer 1T' WTe$_2$ and identified external electric field as an effective tuning knob for its spin lifetime. 
%
%\textcolor{blue}{instead of `what is the \ldots', explicitly spell out `we show that the dominant spin relaxation mechanism is \ldots'}, and how it responds to external electric field as an effective tuning knob. 
%This work provides important insights on understanding dissipation processes in topological systems. 
%
%\textcolor{red}{Here I would rephrase as follows: Our calculations help identifying the mechanisms responsible for the spin relaxation dynamics, and how this dynamics is affected by the application of an external electric field\ldots}

\begin{comment}
$l_{s}$ are obtained using the commonly used relation\cite{vzutic2004spintronics,wu2010spin}
$l_{s}=\sqrt{D\tau_{s}}$ based on the drift-diffusion model with
$D$ the diffusion coefficient calculated from first principles (see
details in ``Methods'' section). Our approach is free from empirical
parameters and is thus of great predictive power. Moreover, the inclusion
of the e-ph scattering allows us to predict $\tau_{s}$ and $l_{s}$
at finite temperatures, which is important for device applications.
\textcolor{red}{More points: Spin-momentum locking and lifetime anisotropy,
which is important for spintronics device control; electric field
effect on spin relaxation as a tuning knob. In general, Understanding
on spin relaxation in topological systems is important in the field. }
\end{comment}

\section*{Results and discussions}

\subsection*{Band structures and spin textures}

\begin{figure*}
\includegraphics[scale=0.55]{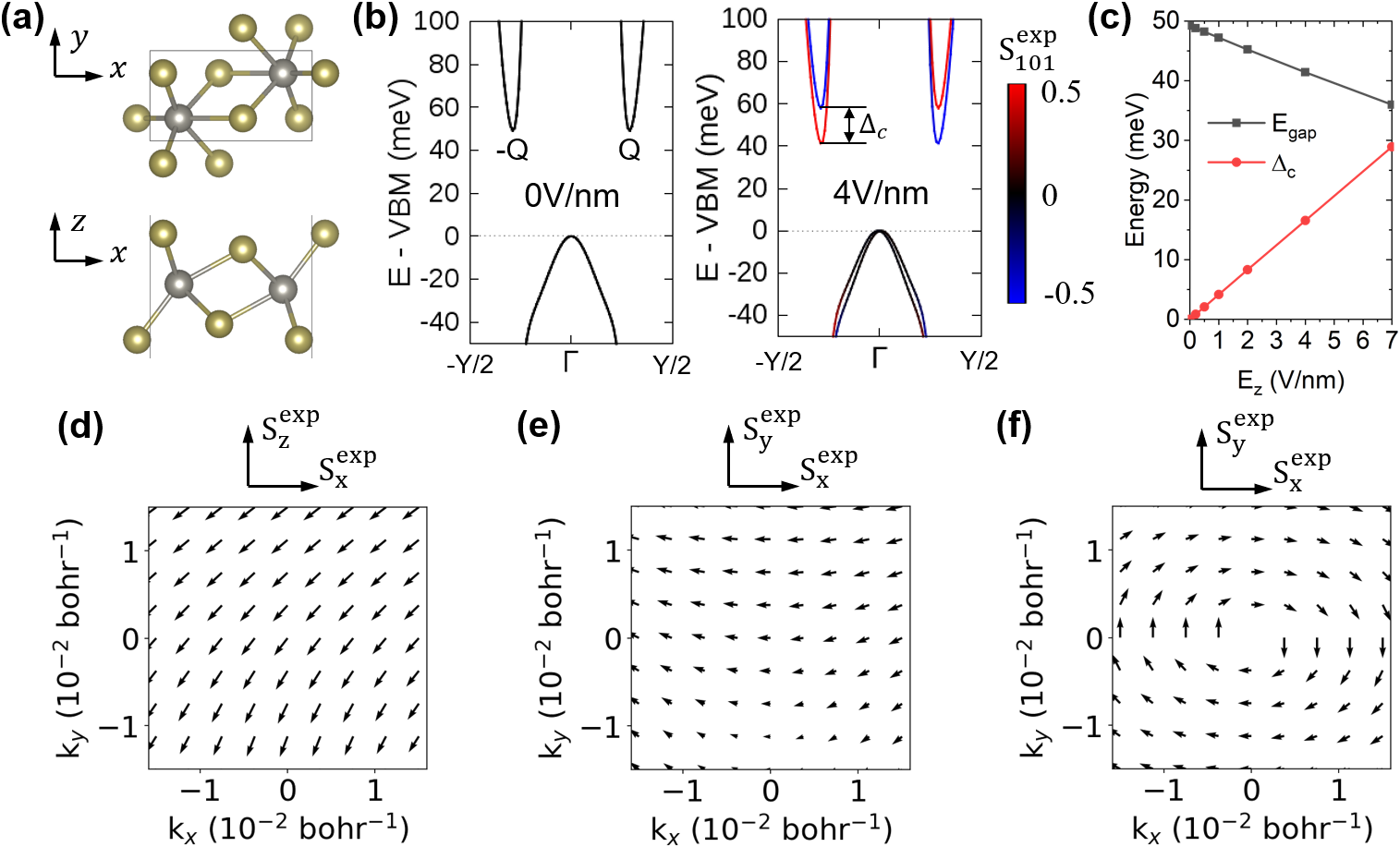}

\caption{Structure, band structures and spin textures of ML-WTe$_{2}$ (a)
The top-view and side-view of the unit cell. (b) Band structures at
$E_{z}=0$ and $E_{z}$=4 V/nm. The color scales the component of
the spin vector ${\bf S}^{\mathrm{exp}}$ projected to the $\left(1,0,1\right)$
direction. ${\bf S}^{\mathrm{exp}}\equiv\left(S_{x}^{\mathrm{exp}},S_{y}^{\mathrm{exp}},S_{z}^{\mathrm{exp}}\right)$
with $S_{i}$ being spin expectation value along direction $i$ and
is the diagonal element of spin matrix $s_{i}$ in Bloch basis. (c)
Band gap $E_{g}$ and conduction band energy splitting $\Delta_{c}$
at ${\bf Q}$ as a function of $E_{z}$. ${\bf Q}$ is the k-point
where conduction band minimum locates and ${\bf Q}\approx\left(0,0.144,0\right)$
in crystal coordinates. (d) and (e) show spin textures at $E_{z}$=4
V/nm centered at ${\bf Q}$ in the $k_{x}$-$k_{y}$ plane of the
lowest conduction band for the plane components of ${\bf S}^{\mathrm{exp}}$
projected into the $S_{x}^{\mathrm{exp}}-S_{z}^{\mathrm{exp}}$ and
$S_{x}^{\mathrm{exp}}-S_{y}^{\mathrm{exp}}$ planes respectively.
(f) is the same as (e) but shows spin textures centered at ${\bf \Gamma}$
of the highest valence band. In subplots (d), (e) and (f), the arrow
length scales the vector length of the plane-projected ${\bf S}^{\mathrm{exp}}$.\label{fig:electronic}}
\end{figure*}

We begin with band structure and spin texture of monolayer(ML)-WTe$_{2}$, which are essential
for understanding spin relaxation mechanism. %Since applying a perpendicular electric field $E_{z}$ is a powerful technique to tune the properties of two-dimensional materials, we examine the effects of $E_{z}$.
The most stable structural phase of ML-WTe$_{2}$ is 1T'~\cite{li2020quantum},
which has inversion symmetry as shown in Fig. \ref{fig:electronic}a.
This corresponds to the QSHI phase below transition temperature, and we will not specifically model the phase transition here.  
%\textcolor{blue}{Clarify that this is the topological phase of interest below the transition temperature, and perhaps mention that we are not explicitly modeling the phase transition.}
Therefore, due to time-reversal and inversion symmetries, every two bands form a Kramers degenerate pair.
A finite perpendicular electric field $E_{z}$ will break the inversion symmetry and induce a k-dependent
internal magnetic field (${\bf B}^{\mathrm{in}}$)\citep{vzutic2004spintronics}.
${\bf B}^{\mathrm{in}}$ split the Kramers pairs and polarize the
spins along different directions (along k-dependent ${\bf B}^{\mathrm{in}}$).

Previous experiments
show that ML-WTe$_{2}$ (1T') is a topological insulator with an indirect
band gap $E_{g}$ of 40-60 meV~\cite{tang2017quantum,maximenko2022nanoscale}. Recent theoretical
study\cite{marrazzo2022thermal} further confirmed the weak temperature renormalization on
the band gap. 
Therefore, the band structure's temperature dependence is not considered in this work. Using the DFT+U approach (see
details in ``Methods'' section), we obtain an $E_{g}$ 49 meV, as
shown in Fig. \ref{fig:electronic}b (left panel). To understand the effect of symmetry broken, we apply an
out-of-plane electric field $E_{z}$, which resulted in band splits at both
conduction and valence bands as shown in Fig.~\ref{fig:electronic}b
(right panel). The band gap $E_{g}$ and spin split at the conduction band minimum $\Delta_{c}$ linearly change with $E_{z}$ (Fig. \ref{fig:electronic}c), with values of 41 and 17
meV respectively, at 4 V/nm.

We next show the spin polarization at band edges under finite electric field ($E=$4 V/nm) in Fig.~\ref{fig:electronic}d-f.  
At conduction band minimum (around $\bf{Q}$) spin polarization is approximately parallel/antiparallel along the $\pm\left(1,0,1\right)$
direction as a ``canted" spin texture (Fig. \ref{fig:electronic}d). Specifically, larger spin expectation values $S^{\mathrm{exp}}$ are observed in xz plane but smaller along y ($S_{y}^{\mathrm{exp}}$). Note that the spins in $-Q$
valley are opposite to those in $Q$ valley. 
Define $\theta_{xz}$
as the angle between the $xz$-plane-projected ${\bf S}^{\mathrm{exp}}$
and the $x$ axis. Around the band edge, $\theta_{xz}$
ranges from 35$^{\circ}$ to 70$^{\circ}$ and the thermally averaged
value at 20-300K ($\left\langle \theta_{xz}\right\rangle $, defined in Eq.~\ref{eq:thermal_average}) is $\sim50^{\circ}$,
in agreement with previous studies\cite{garcia2020canted,zhao2021determination}. For convenience of discussions, we introduce a new definition - canted spin
axis ${\bf S}_{c}^{\mathrm{exp}}$. ${\bf S}_{c}^{\mathrm{exp}}$
is the thermally averaged spin axis, and defined in the caption
of Fig.~\ref{fig:lifetimes}. Based on the discussion above, ${\bf S}_{c}^{\mathrm{exp}}$ is in the $xz$ plane, away from (1,0,1) direction by a few degree. 

The spin polarization at valence band maximum, on the other hand, is mostly circular around $\bf{\Gamma}$ (Fig. \ref{fig:electronic}e), entirely perpendicular to the $\bf{k}$ vector direction, with a Rashba-like spin texture. 
The drastically different spin texture at conduction and valence band edges imply disparate spin relaxation mechanism between electrons and holes in WTe$_2$. We will focus on spin relaxation in conduction electrons below given its interesting canted spin helix texture.

\subsection*{Spin relaxation and its anisotropy}

\begin{figure*}
\includegraphics[scale=0.5]{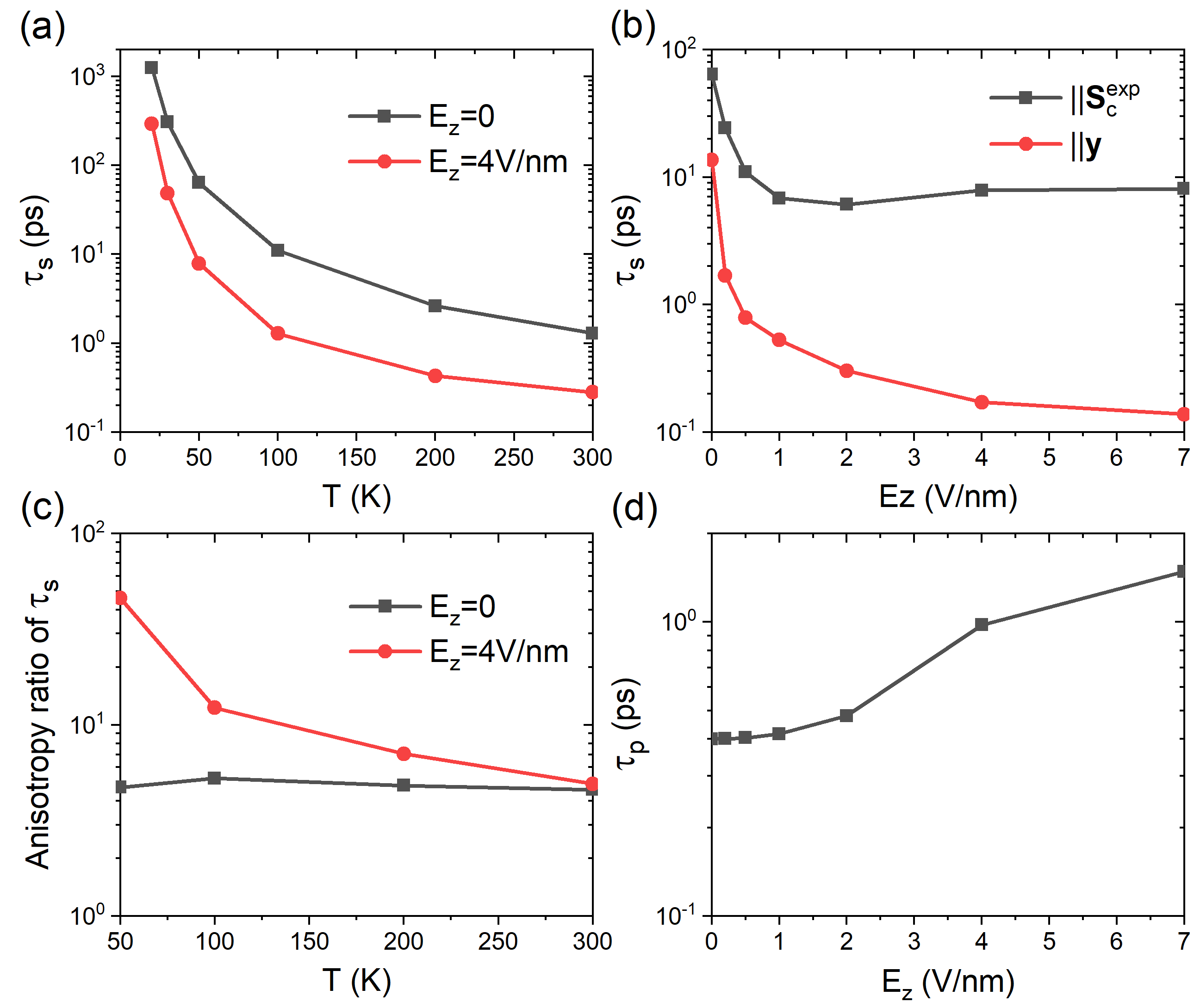}

\caption{Spin and carrier relaxation due to the e-ph scattering of conduction
electrons of ML-WTe$_{2}$. (a) Spin lifetime $\tau_{s}$ along the
canted spin axis ${\bf S}_{c}^{\mathrm{exp}}$ - $\tau_{s}$ as a
function of $T$ at $E_{z}=0$ and 4 V/nm. ${\bf S}_{c}^{\mathrm{exp}}$
is a little away from the $\left(1,0,1\right)$ direction (by a few
degree). 
For conduction electrons, ${\bf S}_{c}^{\mathrm{exp}}$ is
the (thermally) averaged value of the state-resolved vector, which is defined in Eq. \ref{eq:thermal_average} for ${\bf S}^{\mathrm{exp}}\equiv\left(S_{x}^{\mathrm{exp}},S_{y}^{\mathrm{exp}},S_{z}^{\mathrm{exp}}\right)$, i.e. $\left\langle {\bf S}^{\mathrm{exp}}\right\rangle $, for half of BZ closer to $\bf{Q}$ than $\bf{Q'}$. 
%\textcolor{red}{YP: I don't understand why this is necessary: ${\bf S}_{c}^{\mathrm{exp}}$ is
%the (thermally) averaged value of the state-resolved vector,
%$\widetilde{{\bf S}^{\mathrm{exp}}}$,
%i.e., $\left\langle \widetilde{{\bf S}^{\mathrm{exp}}}\right\rangle $.
%$\widetilde{{\bf S}^{\mathrm{exp}}}$ is the spin vector ${\bf S}^{\mathrm{exp}}$
%(defined above) if ${\bf S}^{\mathrm{exp}}$ is around the $\left(1,0,1\right)$
%direction and is $-{\bf S}^{\mathrm{exp}}$ if ${\bf S}^{\mathrm{exp}}$
%is around $-\left(1,0,1\right)$. $\left\langle \widetilde{{\bf S}^{\mathrm{exp}}}\right\rangle $
%is computed using Eq. \ref{eq:thermal_average} }
%
(b) $\tau_{s}$ along
${\bf S}_{c}^{\mathrm{exp}}$ and $\tau_{s}$ along the $y$ axis
($\tau_{s,y}$), approximately perpendicular to the direction of ${\bf S}_{c}^{\mathrm{exp}}$, at 50 K as a function of $E_{z}$. $\tau_{s}$ along
these two directions are the longest and shortest ones respectively.
(c) The anisotropy ratio of $\tau_{s}$, which is the ratio of $\tau_{s}$
along ${\bf S}_{c}^{\mathrm{exp}}$ to along y, as a function
of $T$ at $E_{z}=0$ and 4 V/nm. (d) Carrier lifetime $\tau_{p}$
at 50 K as a function of $E_{z}$.\label{fig:lifetimes}}
\end{figure*}

\begin{comment}
Before examining spin relaxation, we first introduce the canted spin
axis ${\bf S}_{c}^{\mathrm{exp}}$. ${\bf S}_{c}^{\mathrm{exp}}$
is the thermally averaged spin axis and is defined in the caption
of Fig. \ref{fig:lifetimes}.
\textcolor{blue}{May be clearer to introduce this in the discussion of spin texture in previous section.}
${\bf S}_{c}^{\mathrm{exp}}$ is found
in the $xz$ plane and away from the $\left(1,0,1\right)$ direction
by a few degree. Hereinafter, unless specified, $\tau_{s}$ means
$\tau_{s}$ along ${\bf S}_{c}^{\mathrm{exp}}$ - $\tau_{s,c}$.\textcolor{red}{YP: not quite connected here; the description needs to be more direct.}
\end{comment}

Given spin relaxation has the largest spin lifetime anisotropy between along and perpendicular to spin texture directions (during the rotation of principles axis), we will mostly focus on spin relaxation along the spin texture direction (i.e. along canted spin
axis ${\bf S}_{c}^{\mathrm{exp}}$) as $\tau_{s,c}$. 
From Fig.~\ref{fig:lifetimes}a, we find that $\tau_{s,c}$ due to
the e-ph scattering of conduction electrons 
increase with decreasing temperature $T$, at both $E_{z}$=0 and 4 V/nm. Detailed mechanistic discussion will be provided later; briefly, this is a consequence of smaller phonon occupation and weaker electron phonon scattering at lower temperature.    

\begin{comment}
%EY mechanism should be discussed later. 
This is expected because: As discussed below, at 0 and 4 V/nm, spin relaxation is caused by spin-flip e-ph scattering, i.e., EY mechanism. With decreasing $T$, the phonon occupation decreases, so that spin-flip scattering strength is weaken, then spin relaxation becomes slower, i.e., $\tau_{s}$ increases.\textcolor{red}{this paragraph is misplaced}
\end{comment}

In Fig.~\ref{fig:lifetimes}a, we also find that $\tau_{s}$ at 4 V/nm are shorter than $\tau_{s}$
at 0 V/nm. This is attributed to relatively large variation of ${\bf S}^{\mathrm{exp}}$
around ${\bf S}_{c}^{\mathrm{exp}}$ at $E_{z}\neq0$, as shown in Fig.~\ref{fig:electronic}d.
Such variation generally leads to in-homogeneous internal magnetic field and enhances spin relaxation of bulk states thus
reduces $\tau_{s}$, similarly discussed in  Refs~\cite{xu2022substrate,xu2022spin,vzutic2004spintronics,wu2010spin}.
%\textcolor{red}{not quite needed here; you see Ez=0 higher in E dependent plot.}

Moreover, $\tau_{s}$ at 300 K is found short - 0.2-2 ps, in agreement
with experimental data from Refs. \citenum{chen2020anisotropic,lee2019spin,chen2022observation}.
This is consistent with the large spin-mixing parameter $\left\langle b^{2}\right\rangle $
of WTe$_{2}$, which is induced by its large intrinsic SOC strength, i.e. $\sim$0.03
along ${\bf S}_{c}^{\mathrm{exp}}$. 0.03 is much greater than $\left\langle b^{2}\right\rangle $
of light-element materials, e.g., $\left\langle b^{2}\right\rangle $
of conduction electrons of silicon is of $\sim10^{-6}$. Since EY
spin lifetime $\tau_{s}^{\mathrm{EY}}$ usually sets the upper limit
of $\tau_{s}$ and we roughly have $\left(\tau_{s}^{\mathrm{EY}}\right)^{-1}\propto\left\langle b^{2}\right\rangle $,
it is not surprising that $\tau_{s}$ of WTe$_{2}$ is much shorter
than $\tau_{s}$ of silicon, 7 ns at 300 K. Our results suggest that
the long lifetimes ($\gtrsim$100 ps) of WTe$_{2}$ at 300 K observed
experimentally\cite{seifert2019plane,wang2018room} are not $\tau_{s}$
of bulk electron or hole carriers but other sources, such as lifetimes of electrons in edge
states, excitons, etc.

From Fig. \ref{fig:lifetimes}b, we find that $\tau_{s}$ along spin helix direction (${\bf S}_{c}^{\mathrm{exp}}$) and 
along the $y$ axis (approximate perpendicular direction to ${\bf S}_{c}^{\mathrm{exp}}$) at 50 K both decrease with $E_{z}$.
But compared with $\tau_{s,c}$, $\tau_{s,y}$ is much shorter and has stronger dependence on $E_{z}$. At
$E_{z}=0$, the system has inversion symmetry and spin degeneracy, where spin mixing by SOC ($\left\langle b^{2}\right\rangle$) causes spin relaxation. We found $\left\langle b^{2}\right\rangle $ along
$y$ axis is $\sim0.13$, larger than $\left\langle b^{2}\right\rangle $
along ${\bf S}_{c}^{\mathrm{exp}}$, which explains shorter $\tau_{s,y}$. At $E_{z}\neq0$, the inversion symmetry is broken and k-dependent internal magnetic field (${\bf B}^{\mathrm{in}}$) is present. 
%the reason why $\tau_{s,y}\ll\tau_{s,c}$ is: Since bands are splitted by ${\bf B}^{\mathrm{in}}$ and 
%
Given ${\bf S}^{\mathrm{exp}}$ are approximately in $xz$ plane (approximately along the [010] direction), spin
relaxation along $y$ axis is mainly determined by
the fluctuation amplitude of ${\bf B}^{\mathrm{in}}$ projected on the $xz$-plane
- $\Delta{\bf B}_{xz-\mathrm{plane}}^{\mathrm{in}}$~\cite{xu2022substrate,xu2022spin,vzutic2004spintronics,wu2010spin}. Since ${\bf B}^{\mathrm{in}}\equiv2\Delta_{c}{\bf S}^{\mathrm{exp}}/\left(\mu_{B}g_{e}\right)$,
with $\mu_{B}g_{e}$ the electron spin gyromagnetic ratio, $\Delta{\bf B}_{xz-\mathrm{plane}}^{\mathrm{in}}$
is of the order of magnitude of $\Delta_{c}/\left(\mu_{B}g_{e}\right)$, which increases with increasing electric field $E_z$ (Fig.~\ref{fig:electronic}(b)). As a result, $\tau_{s,y}$ monotonically decreases with increasing $E_z$, as  $\Delta{\bf B}_{xz-\mathrm{plane}}^{\mathrm{in}}\propto\Delta_{c}\propto E_{z}$ in Fig.~\ref{fig:lifetimes}b.  On the contrast, spin
relaxation along spin axis ${\bf S}_{c}^{\mathrm{exp}}$ is rather unchanged with increasing E field after 1V/nm, which implies a different spin relaxation mechanism. As will be explained later, it is more dominant by spin-flip scattering processes with little precession, similar to the case of out-of-plane spin relaxation in germenane under E field ~\cite{xu2021giant}. 

%where spins are pinned along the spin canting direction once internal field ${\bf B}^{\mathrm{in}}$ magnitude $\Delta_{c}/\left(\mu_{B}g_{e}\right)$ is big enough.  

%which is large. Therefore, $\tau_{s,y}$ is rather short compared
%with $\tau_{s,c}$. As $\Delta{\bf B}_{xz-\mathrm{plane}}^{\mathrm{in}}\propto\Delta_{c}\propto E_{z}$,
%$\tau_{s,y}$ decrease fast with $E_{z}$.

In Fig. \ref{fig:lifetimes}c, we further show the ratio $\tau_{s,c}/\tau_{s,y}$.
It is found that $\tau_{s,c}/\tau_{s,y}$ at $E_{z}\neq0$ is larger
than that at $E_{z}=0$; and  $\tau_{s,c}/\tau_{s,y}$ at $E_{z}\neq0$
increases rapidly with decreasing $T$. This qualitative trend is similar to the case of spin-momentum locking at low temperature (only possible when persistent spin helix present under $E_{z}$). We realized spin-momentum locking for high SOC Dirac materials under E field in our previous work~\citep{xu2021giant}, where spins parallel to the spin helix direction are dominantly relaxed through intervalley electron-phonon scatterings. But perpendicular to such direction, spins rapidly precess and have a short spin lifetime. Such spin-momentum locking would not maintain at high temperature where high energy phonons are activated (also more electronic states with different spin polarization away from band minimum are occupied), intervalley processes are no longer dominant as intravalley processes are activated. 
Therefore spin lifetime anisotropy is much larger at low temperature compared to high temperature in presence of an external electric field~\cite{xu2021giant, xu2022substrate}.

\begin{comment}
At low $T$, spin relaxation is determined by the states very close
to $\pm{\bf Q}$, as those states are highly polarized along similar
directions, the anisotropy of ${\bf S}^{\mathrm{exp}}$ is huge. Considering
that at $E_{z}\neq0$, $\tau_{s,c}/\tau_{s,y}$ is determined by the
anisotropy of ${\bf S}^{\mathrm{exp}}$, $\tau_{s,c}/\tau_{s,y}$
is therefore huge at low $T$.
\end{comment}

\begin{comment}
Additionally, we show carrier/particle lifetime $\tau_{p}$ as a function
of $E_{z}$ at 50 K in Fig. \ref{fig:lifetimes}d. $\tau_{p}$ is
found increased by several times with $E_{z}$ increasing from 0 to
7 V/nm. We discuss this in the next subsection.
\textcolor{blue}{This is awkward: move the Fig 2d discussion from end of next paragraph here, as it will fit better.}
\end{comment}

\subsection*{Mechanisms of spin relaxation and analytical models}

\begin{figure*}
\includegraphics[scale=0.8]{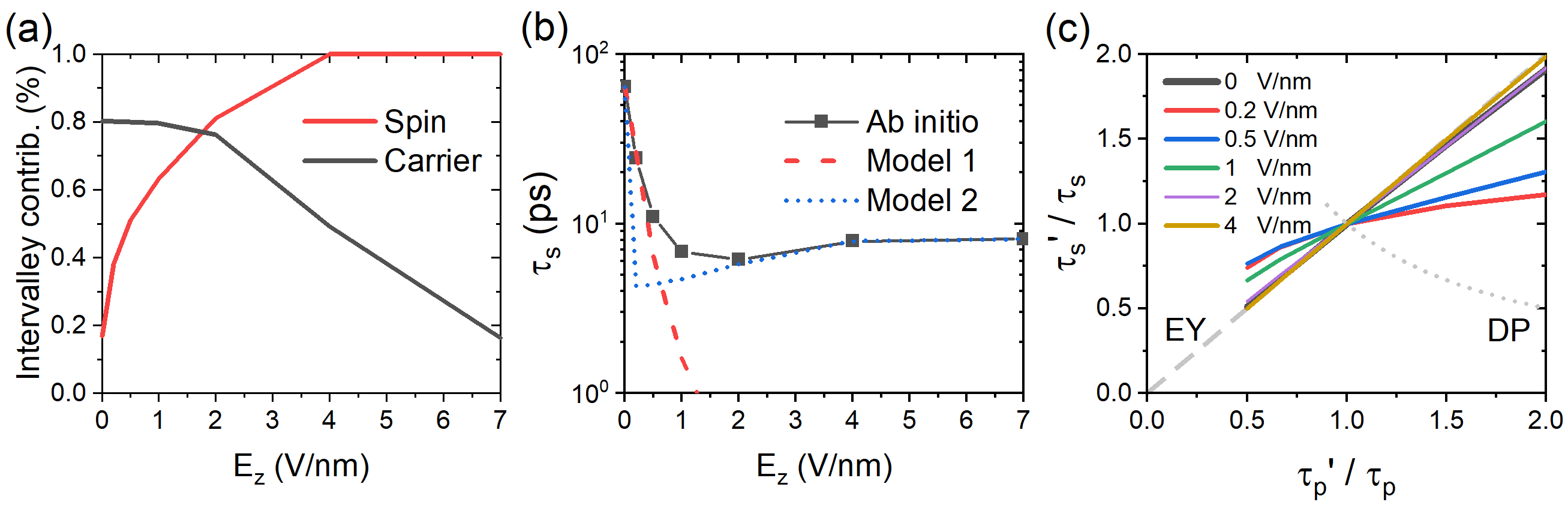}

\caption{Spin and carrier relaxation mechanisms for conduction electrons at
50 K. (a) The relative contribution of intervalley scattering processes
tor spin and carrier relaxation. The intervalley and intravalley scattering
processes correspond to those between $Q$ and $-Q$ valleys and those
within one valley respectively. (b) The FPDM results of $\tau_{s}$
at different $E_{z}$ compared with $\tau_{s}$ obtained by two models.
``Model 1'' is described in the main text. ``Model 2'' is obtained
using Eq. \ref{Eq.semiclassical} - the semiclassical master equation
of the nonequilibrium occupation (the diagonal part of the density
matrix). The results of ``Model 2'' is quite similar to $\tau_{s}$
obtained from Fermi's golden rule considering only spin-flip scattering.
(c) shows the relation between $\tau_{s}'/\tau_{s}$ and $\tau_{p}'/\tau_{p}$
at different $E_{z}$. $\tau_{s}'$ and $\tau_{p}'$ are spin and
carrier lifetimes that are tuned together from their original values
$\tau_{s}$ and $\tau_{p}$ by arbitrarily tuning the scattering strength.
This is realized by multiplying a constant and the generalized (e-ph)
scattering-rate matrix $P$ (Eq. \ref{eq:P}), or equivalently multiplying
a constant and the scattering term of the density-matrix master equation
(Eq. \ref{eq:master}). The EY and DP curves show the ideal relations
in the cases where spin relaxation is purely determined by EY and
DP mechanisms respectively.\label{fig:mechanism}}
\end{figure*}

Since the conduction electrons are in $\pm \bf{Q}$ valleys, spin and carrier
relaxation processes arise from intervalley and intravalley e-ph scatterings.
From Fig. \ref{fig:mechanism}a, we find that intervalley and intravalley
processes play opposite roles in spin and carrier relaxation under E field. At
$E_{z}=0$, intravalley/intervalley processes dominate spin/carrier
relaxation, respectively, while at $E_{z}\geq$4 V/nm, intervalley/intravalley processes
dominate spin/carrier relaxation. 
Continuing the spin-momentum locking discussion earlier, the high-$E_{z}$ behaviour is more detailed as below. 
As shown in Fig. \ref{fig:electronic}b, at large $E_{z}$,
bands are splitted to spin-up and spin-down ones (along ${\bf S}_{c}^{\mathrm{exp}}$)
and the splitting energies are large $\sim$29 meV. So only high-energy 
phonons can contribute to the e-ph scattering between two split bands;
however, the occupation of the phonon bands is low, especially at low $T$,
e.g., 50 K. Therefore, the e-ph scattering processes between two splitted
bands within one valley are weak at large $E_{z}$ and low $T$, and the intervalley scattering process is more dominant. 
On the other hand, for carrier relaxation, at large
$E_{z}$ and low $T$, the intervalley interband spin-conserving e-ph scattering
becomes weak, while the intravalley intraband spin-conserving scattering
is only slightly changed (by 20$\%$). Therefore, at large $E_{z}$,
intravalley scattering dominates carrier relaxation and $\tau_{p}$
becomes long at such condition as shown in Fig. \ref{fig:lifetimes}d. 
%Similarly, at large $E_{z}$, since intravalley interband spin-flip scattering becomes weak intervalley scattering dominates spin relaxation.}

To gain more in-depth physical insights, we then analyze the spin relaxation mechanisms using simplified analytical models.
In
general, with inversion symmetry, bands are Kramers degenerate so
that spin-up and spin-down along an arbitrary axis are well defined
by diagonalizing the corresponding spin matrices. Therefore,
spin relaxation should be Elliot-Yafet(EY) type at $E_{z}=0$. At $E_{z}\neq0$,
spin relaxation is modified by k-dependent ${\bf B}^{\mathrm{in}}$, which has non-trivial consequences.
%Then, the key question is how the effects of ${\bf B}^{\mathrm{in}}$ on spin relaxation in ML-WTe$_{2}$ are.

In the case that ${\bf B}^{\mathrm{in}}$ at all k-points is oriented along
the same axis (e.g. persistent spin helix), spin relaxation along this axis is still EY type, and
the only effect of ${\bf B}^{\mathrm{in}}$ is through the energy
conservation conditions of the e-ph scattering processes with a change of
the electron energies. In more general cases, ${\bf B}^{\mathrm{in}}$
at different k-points is oriented along different axes, its influence on the spin dynamics can be understood by means of the following two
Models:

\textbf{Model 1: Precession-induced spin relaxation.} When the excess
or excited spins are not exactly along ${\bf B}^{\mathrm{in}}$, they
precess about ${\bf B}^{\mathrm{in}}$. The spin precession opens
a spin relaxation channel additional to the existing EY channel without
${\bf B}^{\mathrm{in}}$. Thus, $\tau_{s}$ is approximated as

\begin{align}
\tau_{s}^{-1}\left(E_{z}\right)= & \tau_{s}^{-1}\left(E_{z}=0\right)+\left(\tau_{s}^{-1}\right)^{\Delta{\bf B}_{\perp}^{\mathrm{in}}}\left(E_{z}\right),\label{eq:model1.1}
\end{align}

where $\left(\tau_{s}^{-1}\right)^{\Delta{\bf B}_{\perp}^{\mathrm{in}}}\left(E_{z}\right)$
is the precession term of spin relaxation rate (more accurately, coexistence of precession and scattering). $\left(\tau_{s}^{-1}\right)^{\Delta{\bf B}_{\perp}^{\mathrm{in}}}\left(E_{z}\right)$
is proportional to the fluctuation amplitude $\Delta{\bf B}_{\perp}^{\mathrm{in}}$ of the component ${\bf B}_{\perp}^{\mathrm{in}}$ of the field perpendicular to the local spin direction. Since $\left(\tau_{s}^{-1}\right)^{\Delta{\bf B}_{\perp}^{\mathrm{in}}}\left(E_{z}\right)$
increases with $\Delta{\bf B}_{\perp}^{\mathrm{in}}$ and $\Delta{\bf B}_{\perp}^{\mathrm{in}}\propto E_{z}$,
$\left(\tau_{s}^{-1}\right)^{\Delta{\bf B}_{\perp}^{\mathrm{in}}}\left(E_{z}\right)$
increases with $E_{z}$. In previous theoretical works, $\left(\tau_{s}^{-1}\right)^{\Delta{\bf B}_{\perp}^{\mathrm{in}}}\left(E_{z}\right)$
is often estimated using the DP (Dyakonov Perel) relation\cite{xu2022spin,vzutic2004spintronics,wu2010spin}

\begin{align}
\left(\tau_{s}^{-1}\right)^{\mathrm{DP}}= & \tau_{p}\left[\Delta{\bf B}_{\perp}^{\mathrm{in}}/\left(\mu_{B}g_{e}\right)\right]^{2}.\label{eq:model1.2}
\end{align}

\textbf{Model 2: Spin splitting and enhancement of spin-flip matrix
elements.} If ${\bf B}^{\mathrm{in}}\equiv2\Delta_{c}{\bf S}^{\mathrm{exp}}/\left(\mu_{B}g_{e}\right)$
are nonzero and ${\bf S}^{\mathrm{exp}}$ at different k-points are
approximately along a ``canted'' axis ${\bf S}_{c}^{\mathrm{exp}}$,
spin precession may be unimportant to spin relaxation along ${\bf S}_{c}^{\mathrm{exp}}$,
so that spin relaxation is dominated by EY mechanism. In this case,
spin relaxation rate is well approximated by 
the spin-flip e-ph scattering with the Fermi's Golden rule or the semiclassical
form of the density-matrix master equation (Eq. \ref{Eq.semiclassical}).

However, since the variation of ${\bf S}^{\mathrm{exp}}$ is not negligible
now, ${\bf B}^{\mathrm{in}}$ not only affect the energy conservation
conditions but also enhance the spin-flip e-ph matrix elements\cite{xu2022substrate}.
According to the Fermi's golden rule, the enhancement of spin-flip matrix
elements enhances $\tau_{s}^{-1}$ and reduces $\tau_{s}$.

In Fig. \ref{fig:mechanism}b, we compare the FPDM results of $\tau_{s}$
with $\tau_{s}$ obtained using the above two models. We find that
``Model 1'' well describes the decrease of $\tau_{s}$ with $E_{z}$
at low $E_{z}$ but underestimate $\tau_{s}$ at $E_{z}$\textgreater
0.5 V/nm. ``Model 2'' describes well the cases at $E_{z}=0$ and
$E_{z}\geq$2 V/nm. The comparison indicates that: First, at $E_{z}=0$
and high $E_{z}$ ($\geq$2 V/nm), spin relaxation is EY type and
spin precession indeed plays a negligible role. Second, at low but nonzero $E_{z}$
(0\textless$E_{z}$\textless 2 V/nm), spin precession is important but its effects are not simply described by the
model relations, Eqs. \ref{eq:model1.1} or \ref{eq:model1.2}.

We further examine spin relaxation mechanism from another point of
view. Conventionally, spin relaxation mechanism is determined from
the relation between $\tau_{s}$ and $\tau_{p}$ - EY mechanism leads
to $\tau_{s}\propto\tau_{p}$ while DP mechanism leads to $\tau_{s}\propto\tau_{p}^{-1}$.
Theoretically, this proportionality can be evaluated by introducing a factor $F^{\mathrm{sc}}$
to scale the scattering strength. Computationally, this is realized by
multiplying $F^{\mathrm{sc}}$ to the scattering term of the density-matrix
master equation (the second term of Eq. \ref{eq:master}) but keep
the coherent term (the first term of Eq. \ref{eq:master}) unchanged.
This is equivalent to multiply $F^{\mathrm{sc}}$ to all elements
of the generalized scattering-rate matrix $P$ (Eq. \ref{eq:P}).
To avoid confusions , we name carrier and spin lifetimes after introducing
$F^{\mathrm{sc}}$ as $\tau_{s}'$ and $\tau_{p}'$ respectively.
Note that when $F^{\mathrm{sc}}=1$,  $\tau_{s}'\equiv\tau_{s}$ and
$\tau_{p}'\equiv\tau_{p}$. According to Eq. \ref{eq:carrier_lifetime},
we have $\tau_{p}'=\left(F^{\mathrm{sc}}\right)^{-1}\tau_{p}$, i.e.,
$\tau_{p}'/\tau_{p}=\left(F^{\mathrm{sc}}\right)^{-1}$. Importantly, $\tau_{s}'/\tau_{s}$ proportionality to $F^{\mathrm{sc}}$ depends on the spin relaxation mechanism: For EY
mechanism, as $\tau_{s}'\propto\tau_{p}'$, 
we can show easily $\tau_{s}'/\tau_{s}=\left(F^{\mathrm{sc}}\right)^{-1}=\tau_{p}'/\tau_{p}$. 
%we can assume $\tau_{s}'=C\tau_{p}'=C\left(F^{\mathrm{sc}}\right)^{-1}\tau_{p}$.
%Since this relation holds when $F^{\mathrm{sc}}=1$, we have $\tau_{s}=C\tau_{p}$.
%So $\tau_{s}'=\left(F^{\mathrm{sc}}\right)^{-1}\tau_{s}$, i.e., $\tau_{s}'/\tau_{s}=\left(F^{\mathrm{sc}}\right)^{-1}=\tau_{p}'/\tau_{p}$.
Similarly, for DP mechanism, we can prove that $\tau_{s}'/\tau_{s}=F^{\mathrm{sc}}=\left(\tau_{p}'/\tau_{p}\right)^{-1}$.
Therefore, the relation between $\tau_{s}'/\tau_{s}$ and $\tau_{p}'/\tau_{p}$
is useful to understand spin relaxation mechanism qualitatively.

From Fig. \ref{fig:mechanism}c, we find that at $E_{z}=0$ and $E_{z}\geq$2
V/nm, $\tau_{s}'/\tau_{s}\approx\tau_{p}'/\tau_{p}$, which indicates
EY mechanism. When 0\textless$E_{z}$\textless 2 V/nm, $\tau_{s}'/\tau_{s}$
increases with $\tau_{p}'/\tau_{p}$ but with a smaller ratio than 1,
which indicates that spin relaxation is neither EY nor DP type in this range. These
findings are consistent with our analytical model discussions related to Fig.~\ref{fig:mechanism}b.

\subsection*{Spin transport and diffusion length}

\begin{figure}
\includegraphics[scale=0.7]{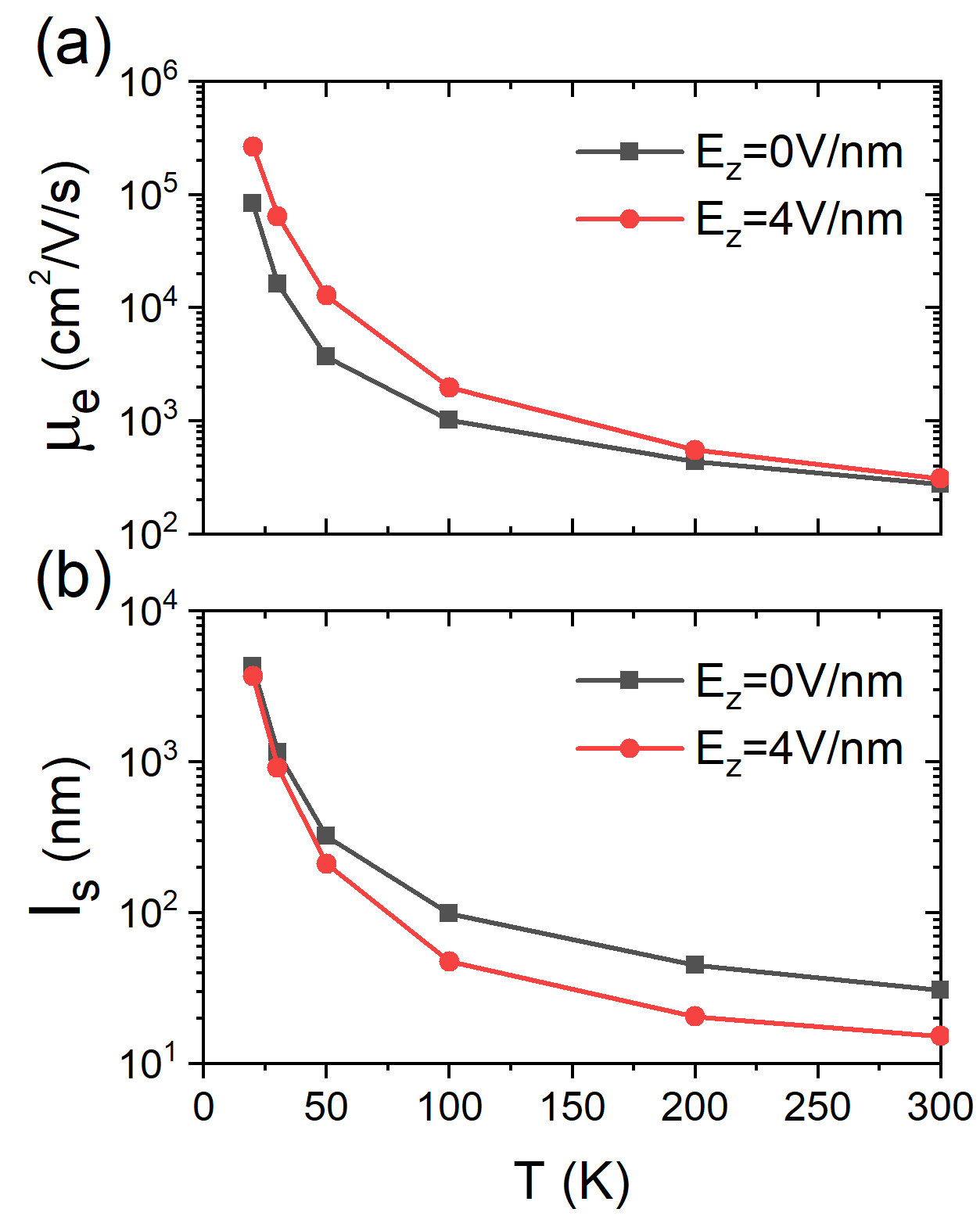}

\caption{The \emph{ab initio} results of (a) the electron mobility $\mu_{e}$
and (b) the conduction electron spin diffusion length $l_{s}$ of
undoped monolayer 1T'-WTe$_{2}$ as a function of $T$ at $E_{z}=0$
and 4 V/nm.\label{fig:transport}}
\end{figure}

At the end, we show our \emph{ab initio} results of carrier and spin
transport properties in Fig. \ref{fig:transport}. Similar to $\tau_{s}$,
the electron mobility $\mu_{e}$ and the spin diffusion length $l_{s}$
increase with decreasing $T$, which is a result of the e-ph scattering
being weaker at lower $T$. $\mu_{e}$ of ML-WTe$_{2}$ are found
relatively high, e.g., 1000-3000 cm$^{2}$/V/s at 100 K. Since $\mu_{e}$
is proportional to $\tau_{p}$ and $\tau_{p}$ increases with $E_{z}$,
we expect that $\mu_{e}$ increases with $E_{z}$. This is confirmed
in Fig. \ref{fig:transport}a. As discussed earlier, $\tau_{s}$ has an opposite trend from $\tau_{p}$, namely, decreases with increasing E field. 
$l_{s}$ is computed approximately based on the drift-diffusion
model\cite{vzutic2004spintronics} and the generalized Einstein relation\citep{kubo1966fluctuation}
(Eq. \ref{eq:D}), with $l_{s}\propto\sqrt{\mu_{e}\tau_{s}}$.
The $E_{z}$ dependence of $l_{s}$ is a net result of $E_{z}$ dependence
of $\mu_{e}$ and $\tau_{s}$, where $\tau_{s}$ dominates the trend. From Fig.~\ref{fig:transport}b, we
find that increasing $E_{z}$ reduces $l_{s}$ significantly at 300
K but affects $l_{s}$ a little at $T\le$100 K.
Our theoretical $l_{s}$ are 10-30 nm at 300 K, in agreement with
previous experimental and theoretical results\cite{zhao2020observation,zhang2017tunable,vila2021low}.
Without defects, $l_{s}$ can reach relatively long at low $T$, e.g., 300 nm
at 50 K. This makes ML-WTe$_{2}$ promising for spintronic applications.

\section*{Conclusions}

Using our FPDM approach, we simulate $\tau_{s}$ and $l_{s}$ of 1T' ML-WTe$_{2}$ at different $T$ and $E_{z}$ including spin-phonon scattering effects. We predict that $\tau_{s}$ and $l_{s}$ of bulk-state
electrons are respectively $\sim$1 ps and $\sim$30 nm at room temperature,
consistent with previous experimental data\cite{chen2020anisotropic,lee2019spin,chen2022observation,zhao2020observation,zhang2017tunable}.
Without considering impurities, the values of $\tau_{s}$ and $l_{s}$ become long
at low $T$, e.g., 100 ps and 300 nm respectively at 50 K. These findings suggest
WTe$_{2}$'s promise for spintronic applications. In addition, at
$E_{z}\neq0$, $\tau_{s}$ is highly anisotropic and the ratio of
the longest $\tau_{s}$ to the shortest value can reach $\sim$60.
The expectation value ${\bf S}^{\mathrm{exp}}$ for conduction electrons is approximately ``canted'' along a particular direction ${\bf S}_{c}^{\mathrm{exp}}$,
which is in the $xz$ plane and has an angle to the $x$ axis of about
50$^{\circ}$. However, the spins are found not strictly ``canted'' and
the variations of ${\bf S}^{\mathrm{exp}}$ of conduction electrons
are non-negligible. Since such variations enhance the spin relaxation
by producing an additional relaxation channel or enhancing the spin-flip
e-ph matrix elements, it is found that increasing $E_{z}$ within a range of $<$1V/nm can significantly
reduce $\tau_{s}$. Based on detailed analysis, we find that at zero
and large $E_{z}$ (e.g., $\ge$2 V/nm for 50 K), spin relaxation
is dominated by EY mechanism, but at low but nonzero $E_{z}$, spin
relaxation is neither EY nor DP type.

\section*{Methods}

\subsection*{First-principles density-matrix dynamics for spin relaxation}

We solve the quantum master equation of density matrix $\rho\left(t\right)$
as the following:\cite{xu2021ab} 
\begin{align}
\frac{d\rho_{12}\left(t\right)}{dt}= & \left[H_{e},\rho\left(t\right)\right]_{12}+\nonumber \\
 & \left(\begin{array}{c}
\frac{1}{2}\sum_{345}\left\{ \begin{array}{c}
\left[I-\rho\left(t\right)\right]_{13}P_{32,45}\rho_{45}\left(t\right)\\
-\left[I-\rho\left(t\right)\right]_{45}P_{45,13}^{*}\rho_{32}\left(t\right)
\end{array}\right\} \\
+H.C.
\end{array}\right),\label{eq:master}
\end{align}
Eq.~\ref{eq:master} is expressed in the Schr\"odinger picture, where
the first and second terms on the right hand side of the equation relate
to the coherent and incoherent dynamics, which can lead to Larmor precession, and
scattering processes respectively. $H_{e}$ is the electronic Hamiltonian.
$\left[H,\rho\right]\,=\,H\rho-\rho H$. H.C. is Hermitian conjugate.
The subindex, e.g., ``1'' is the combined index of $\textbf{k}$-point
and band. The weights of k-points must be considered when doing sum
over k points. $P$ is the generalized scattering-rate matrix for
the e-ph scattering and reads\cite{xu2021ab}

\begin{align}
P_{1234}= & \sum_{q\lambda\pm}A_{13}^{q\lambda\pm}A_{24}^{q\lambda\pm,*},\label{eq:P}\\
A_{13}^{q\lambda\pm}= & \sqrt{\frac{2\pi}{\hbar}}g_{12}^{q\lambda\pm}\sqrt{\delta_{\sigma}^{G}\left(\epsilon_{1}-\epsilon_{2}\pm\omega_{q\lambda}\right)}\sqrt{n_{q\lambda}^{\pm}},
\end{align}
where $q$ and $\lambda$ are phonon wavevector and mode, $g^{q\lambda\pm}$
is the e-ph matrix element, resulting from the absorption ($-$) or
emission ($+$) of a phonon, computed with self-consistent SOC from
first-principles\citep{giustino2017electron}, $n_{q\lambda}^{\pm}=n_{q\lambda}+0.5\pm0.5$
where $n_{q\lambda}$ are the phonon Bose factors, and $\delta_{\sigma}^{G}$
represents an energy conserving $\delta$-function broadened to a
Gaussian of width $\sigma$.

Starting from an initial density matrix $\rho\left(t_{0}\right)$
prepared with a net spin, we evolve $\rho\left(t\right)$ through
Eq. \ref{eq:master} for a long enough time, typically from a few
ps to a few ns. We then obtain spin observable $S\left(t\right)$
from $\rho\left(t\right)$ (Eq. S1) and extract spin lifetime $\tau_{s}$
from $S\left(t\right)$ using Eq. S2.

More details are given in Supporting Information Sec. SI and Ref.
\citenum{xu2021ab}.

\subsection*{Carrier relaxation and transport properties}

In the semiclassical limit, the density matrix $\rho$ is replaced by the
(nonequilibrium) occupation $f$, then the scattering term of Eq.~\ref{eq:master}
becomes\cite{xu2021ab}

\begin{align}
\frac{df_{1}}{dt}= & \mathop{\sum_{2\neq1}}\left[\left(1-f_{1}\right)P_{11,22}f_{2}-\left(1-f_{2}\right)P_{22,11}f_{1}\right].\label{Eq.semiclassical}
\end{align}

From the above equation, we can derive the electronic-state-resolved
carrier relaxation rate\cite{xu2021ab}

\begin{align}
\frac{1}{\tau_{p,1}}= & \mathop{\sum_{2\neq1}}\left[P_{11,22}f_{2}^{\mathrm{eq}}+\left(1-f_{2}^{\mathrm{eq}}\right)P_{22,11}\right],\label{eq:carrier_lifetime}
\end{align}

where $f_{2}^{\mathrm{eq}}$ is the equilibrium occupation of state
``2'' and is Fermi-Dirac function of the energy of state ``2''.

The carrier lifetime $\tau_{p}$ is defined as

\begin{align}
\tau_{p}= & 1/\left\langle \tau_{p}^{-1}\right\rangle ,
\end{align}

where $\left\langle A\right\rangle $ means the thermally averaged
value of electronic-state-resolved quantity $A$ and reads

\begin{align}
\left\langle A\right\rangle = & \frac{\sum_{1}\left(-\left[f^{\mathrm{eq}}\right]'_{1}\right)A_{1}}{\sum_{1}\left(-\left[f^{\mathrm{eq}}\right]'_{1}\right)},\label{eq:thermal_average}
\end{align}

where $\left[f^{\mathrm{eq}}\right]'$ is the derivative of the Fermi-Dirac
distribution function.

We calculate the electron mobility $\mu_{e}$ by solving the linearized
Boltzmann equation in momentum-relaxation-time approximation\citep{ciccarino2018dynamics,gunst2016first,mahan2000many},
\begin{align}
\mu_{e,i}= & \frac{e}{n_{e}V_{u}N_{k}}\sum_{1\in\mathrm{CB}}\left[f^{\mathrm{eq}}\right]'_{1}v_{1,i}^{2}\tau_{m,1},\label{eq:mobility}
\end{align}
where $i=x,y$ for two dimensional systems. $N_{k}$ is the number
of k points. $V_{u}$ is the unit cell volume. $n_{e}$ is electron
density. $\mathrm{CB}$ denotes conduction bands. $\varepsilon_{\mu}$
is the chemical potential. $v$ is the band velocity. $\tau_{m}$
is the momentum relaxation time and is approximately\citep{ciccarino2018dynamics,gunst2016first,mahan2000many}
\begin{align}
\tau_{m,1}^{-1}= & \mathop{\sum_{2\neq1}}\left\{ \left[P_{11,22}f_{2}^{\mathrm{eq}}+\left(1-f_{2}^{\mathrm{eq}}\right)P_{22,11}\right]\mathrm{cos}\theta_{12}\right\} ,\label{eq:cos_theta}\\
\mathrm{cos}\theta_{12}= & \frac{{\bf v}_{1}\cdot{\bf v}_{2}}{v_{1}v_{2}},
\end{align}

where ${\bf v}$ is the velocity vector.

With the computed mobility, the diffusion coefficient $D$ (of conduction
electrons) is computed using the general form of Einstein relation\citep{kubo1966fluctuation}

\begin{align}
D= & -\mu_{e}n_{e}/\frac{dn_{e}}{d\varepsilon_{\mu}}.\label{eq:D}
\end{align}

At low density, since $n_{e}/\frac{dn_{e}}{d\varepsilon_{\mu}}\approx-k_{B}T$,
$D\approx\mu_{e}k_{B}T$.

\subsection*{Computational details}

The ground-state electronic structure, phonons, as well as the e-ph
matrix elements are firstly calculated using density functional theory
(DFT) with relatively coarse $k$ and $q$ meshes in the DFT plane-wave
code JDFTx\citep{sundararaman2017jdftx}. The standard density-functional-theory
(DFT) calculations with Perdew-Burke-Ernzerhof (PBE) or local-density-approximation
(LDA) exchange-correlation functional predicts ML-WTe$_{2}$ to be
a Weyl semimetal while ML-WTe$_{2}$ is found to be a topological insulator
experimentally. To cure this issue, we include
Hubbard corrections~\citep{Dudarev1998-mz} in exchange-correlation functional (PBE+U)~\citep{perdew1996generalized}, which open up the band gap for monolayer WTe$_{2}$.
%The exchange-correlation functional is PBE\citep{perdew1996generalized}.
Hubbard corrections are considered for the $d$ orbitals of W atom
and U is 2.5 eV so that the band gap is 50 meV close to the experimental
value\cite{Tang2017-ja} of 55 meV. 
The lattice constants and internal
geometries are fully relaxed with DFT+U. We use Optimized Norm-Conserving
Vanderbilt (ONCV) pseudopotentials\citep{hamann2013optimized} with
self-consistent SOC throughout, which we find converged at a kinetic
energy cutoff of 64 Ry. The DFT calculations use 4$\times$8 $k$
meshes. The phonon calculation employs a $2\times4$ supercell through
finite difference calculations. The Coulomb truncation technique\citep{ismail2006truncation}
is employed to accelerate convergence with vacuum sizes. The vacuum
size is 30 bohr (additional to the thickness of the heterostructures)
and is found large enough to converge the final results of spin lifetimes.
The electric field along the non-periodic direction is applied as
a ramp or saw-like potential.

We then transform all quantities from plane wave basis to maximally
localized Wannier function basis\citep{marzari1997maximally}, and
interpolate them\citep{PhononAssisted,giustino2017electron,GraphiteHotCarriers,brown2017experimental,NitrideCarriers,TAparameters}
to substantially finer k and q meshes. The fine $k$ and $q$ meshes
are $160\times320$ for simulations at 300 K and are finer at lower
temperature, e.g., $400\times800$ for simulations at 50 K. The Wannier
interpolation approach fully accounts for polar terms in the e-ph
matrix elements and phonon dispersion relations using the method
of Sohier et al.\cite{sohier2017breakdown} for the 2D systems.

The real-time dynamics simulations are performed with the
DMD code interfaced to JDFTx. The energy-conservation smearing parameter
$\sigma$ is chosen to be comparable or smaller than $k_{B}T$ for
each calculation, e.g., 10 meV and 3.3 meV at 300 K and 50 K respectively.

\section*{DATA AVAILABILITY}

The data that support the findings of this study are available upon
request to the corresponding author.

\section*{CODE AVAILABILITY}

The codes that were used in this study are available on github \href{https://github.com/Ping-Group-UCSC/denmat_dynm_development}{DMD-code} and \href{https://github.com/shankar1729/jdftx}{JDFTx}.

\section*{Acknowledgements}

This work is supported by the computational chem-
ical science program within the Office of Science at DOE
under grant No. DE-SC0023301. 
This research used resources of the Center
for Functional Nanomaterials, which is a US DOE Office of Science
Facility, and the Scientific Data and Computing center, a component
of the Computational Science Initiative, at Brookhaven National Laboratory
under Contract No. DE-SC0012704, the National Energy Research
Scientific Computing Center (NERSC) a U.S. Department of Energy Office
of Science User Facility operated under Contract No. DE-AC02-05CH11231.

\section*{Author contributions}

J.X. and H.T. performed the first-principles calculations.  J.X., H.T., R.S., and Y.P. analyzed
the results. Y.P. designed and supervised the study. J.X.
and Y.P. wrote the first draft of the manuscript.
All authors contribute to the manuscript preparation. 
%\textcolor{blue}{Need to add at least something minor for Hiro and me in this explicit statement to justify authorship: maybe include H.T. in performed the first-principles calculations, and R.S. in formulated the methodology or something equivalent.}

\section*{ADDITIONAL INFORMATION}

\textbf{Supplementary Information} is available for this paper at {[}url{]}.

\textbf{Competing interests:} The authors declare no competing interests.

\section*{REFERENCES}

 \bibliographystyle{apsrev4-1}
\bibliography{ref}

\end{document}

\title{Supplemental Material for: Spin relaxation and diffusion in monolayer 1T'-WTe$_2$ from first principles} 
\setcounter{page}{1}   
\date{\today} 
\author{Junqing Xu\footnotemark[1]} 
\thanks{Current Address: Department of Physics, Hefei University of Technology, Hefei, Anhui, China}
\affiliation{Department of Chemistry and Biochemistry, University of California, Santa Cruz, CA 95064, USA}
\author{Hiroyuki Takenaka\footnotemark[1]} 
\thanks{HT and JX contributed equally} 
\email{Deceased: Hiroyuki Takenaka}
\affiliation{Department of Chemistry and Biochemistry, University of California, Santa Cruz, CA 95064, USA} 
\author{Andrew Grieder} %TBD
\affiliation{Department of Materials Science and Engineering, University of Wisconsin-Madison, WI, 53706, USA}  
\author{Jacopo Simoni}
\affiliation{Department of Materials Science and Engineering, University of Wisconsin-Madison, WI, 53706, USA}
\author{Ravishankar Sundararaman}
\email{sundar@rpi.edu}  
\affiliation{Department of Materials Science and Engineering, Rensselaer Polytechnic Institute, 110 8th Street, Troy, New York 12180, USA}  
\author{Yuan Ping}
\email{yping3@wisc.edu}
\affiliation{Department of Materials Science and Engineering, University of Wisconsin-Madison, WI, 53706, USA}
\affiliation{Department of Physics, University of Wisconsin - Madison, WI, 53706, USA}
\affiliation{Department of Chemistry, University of Wisconsin - Madison, WI, 53706, USA}

%\maketitle
{ \let\clearpage\relax \maketitle }

\begin{comment}
\section{TO-DO-LIST}

\begin{itemize}
   %\item computational methods from DFT in detail
   \item complete band structure
   \item complete phonon band structure
   \item carrier lifetime and conductivity calculations - if available
   \item DFT+U vs HSE if we have?
\end{itemize} 
\end{comment}

\section{The simulation of spin lifetime}

Spin lifetime is calculated based on the method developed in Ref.
\citenum{xu2021ab}. To define spin lifetime, we follow the time
evolution of the observable 
\begin{align}
S_{i} & \left(t\right)=\mathrm{Tr}\left(s_{i}\left(t\right)\rho\left(t\right)\right),
\end{align}
where $\rho\left(t\right)$ is the density matrix; $s_{i}$ is spin
Pauli matrix in Bl\"och basis along direction $i$. This time evolution
must start at an initial state (at $t\,=\,t_{0}$) with a net spin
i.e. $\delta\rho(t_{0})=\rho(t_{0})-\rho\super{eq}\neq0$ such that
$\delta S_{i}(t_{0})=S_{i}\left(t_{0}\right)-S_{i}\super{eq}\neq0$,
where ``eq'' corresponds to the final equilibrium state. We evolve
the density matrix through the quantum master equation given in Ref.
\citenum{xu2021ab} (Eq. 5 therein) for a long enough simulation
time, typically from a few ps to a few ns, until the evolution of
$S_{i}\left(t\right)$ can be reliably fitted by 
\begin{align}
S_{i}\left(t\right)-S_{i}^{\mathrm{eq}}= & \left[S_{i}\left(t_{0}\right)-S_{i}^{\mathrm{eq}}\right]exp\left[-\frac{t-t_{0}}{\tau_{s,i}}\right]\times\mathrm{cos}\left[\Omega\left(t-t_{0}\right)+\phi\right].\label{eq:exp_decay}
\end{align}
to extract the relaxation time, $\tau_{s,i}$. Above, $\Omega$ is
oscillation frequency due to Larmor precession. For ML-WTe$_{2}$
conduction electrons, spins are approximately ``canted'' along the
canted spin axis ${\bf S}_{c}^{\mathrm{exp}}$, so that for spin relaxation
along ${\bf S}_{c}^{\mathrm{exp}}$, there is no Larmor precession
and $\Omega$ can be safely set to zero.

In Ref. \citenum{xu2021ab}, we have shown that it is suitable to
generate the initial spin imbalance by applying a test magnetic field
at $t=-\infty$, allowing the system to equilibrate with a net spin
and then turning it off suddenly at $t_{0}$.

\begin{figure}[H]
\centering \includegraphics[scale=0.7]{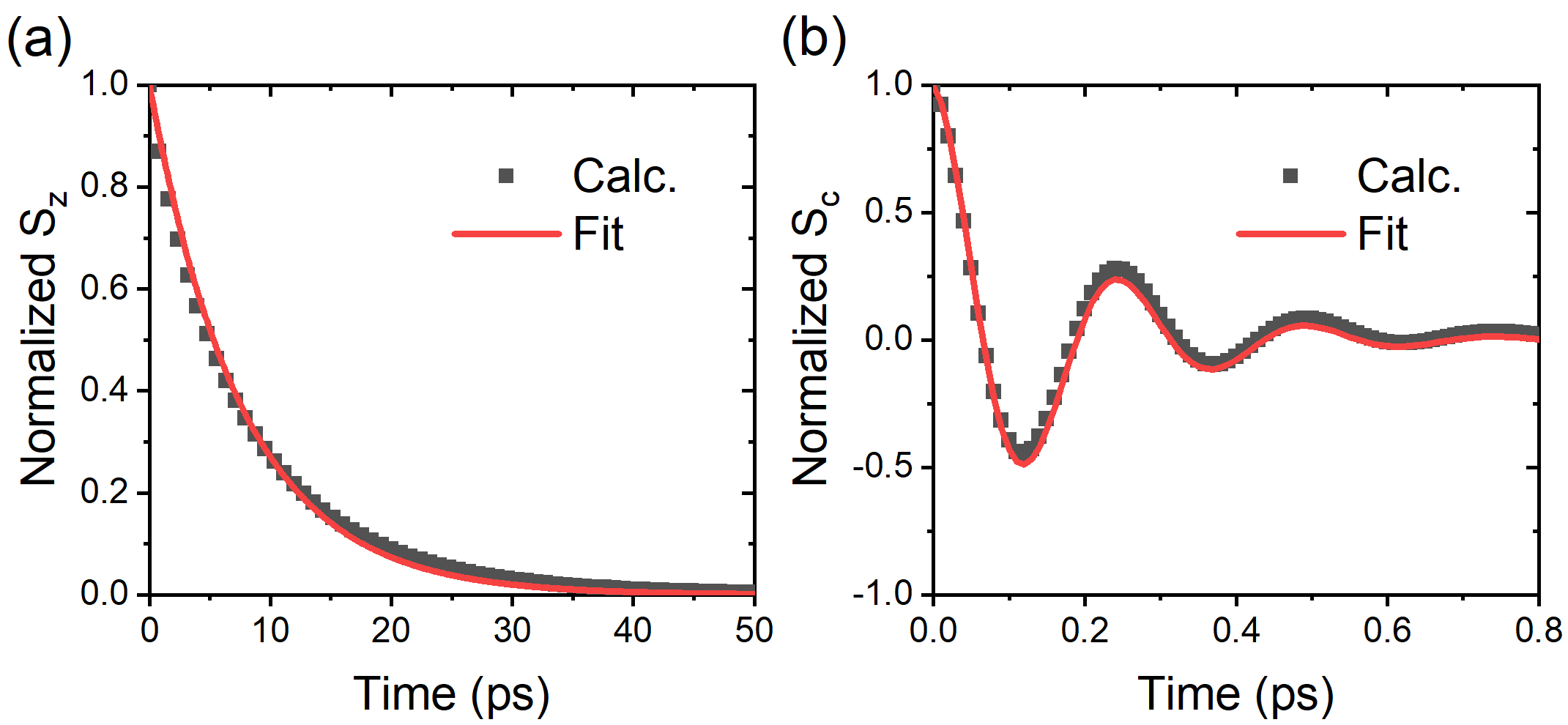} \caption{Time evolutions of (a) $S_{z}$ and (b) spin observable along the
canted spin axis ${\bf S}_{c}^{\mathrm{exp}}$ - $S_{c}$ of conduction
electrons of ML-WTe$_{2}$ at $E_{z}$=7 V/nm and 50 K after the initial
spin imbalance generated by a test magnetic field. ``Calc.'' denotes
calculated results. ``Fit'' denotes fitted results using Eq. \ref{eq:exp_decay}.}
\label{fig:fit}
\end{figure}

In Fig. \ref{fig:fit}, we compare calculated spin observables and
fitted ones using Eq. \ref{eq:exp_decay} of conduction electrons
of ML-WTe$_{2}$ at $E_{z}$=7 V/nm and 50 K after the initial spin
imbalance generated by a test magnetic field. We find the fitted curves
match the calculated ones nicely, which gives spin lifetimes.

\section{Spin mixing parameter}

Suppose the spin of a state ``1'' (which is the combined index of
k-point and band) is highly polarized along $z$ direction. Then in
general, the wavefunction of state ``1'' can be written as $\Psi_{1}\left({\bf r}\right)=a_{z,1}\left({\bf r}\right)\alpha+b_{z,1}\left({\bf r}\right)\beta$,
where $a$ and $b$ are the coefficients of the large and small components
of the wavefunction, and $\alpha$ and $\beta$ are spinors (one up
and one down for direction $z$). Define $a_{z,1}^{2}=\int|a_{z,1}\left({\bf r}\right)|^{2}d{\bf r}$
and $b_{z,1}^{2}=\int|b_{z,1}\left({\bf r}\right)|^{2}d{\bf r}$,
then $a_{z,1}^{2}>b_{z,1}^{2}$ and $b_{z,1}^{2}$ is just spin-mixing
parameter of state ``1'' along direction $z$. With the definition
of spin expectation value $S_{z,1}^{\mathrm{exp}}=s_{z,11}$, which
is the diagonal element for state "1" of spin Pauli matrix along
direction $z$ in Bloch basis, we have
\begin{align}
a_{z,1}^{2}+b_{z,1}^{2}= & 1,\\
0.5\left(a_{z,1}^{2}-b_{z,1}^{2}\right)= & S_{z,1}^{\mathrm{exp}},
\end{align}
Therefore, 
\begin{align}
b_{z,1}^{2}= & 0.5-S_{z,1}^{\mathrm{exp}}.
\end{align}

\bibliographystyle{apsrev4-1}
\bibliography{ref}